\documentclass[12pt]{article}

\newcommand{\be}{\begin{eqnarray}}
\newcommand{\ee}{\end{eqnarray}}
\newcommand{\ca}{C_{\alpha}}
\newcommand{\cb}{C_{\beta}}
\newcommand{\cbd}{C_{\beta}^{\dagger}}
\newcommand{\dab}{d(\alpha,\beta)}
\newcommand{\qp}{q^{\prime}}
\newcommand{\qdp}{q^{\prime \prime}}
\newcommand{\tp}{t^{\prime}}
\newcommand{\tdp}{t^{\prime \prime}}
\newcommand{\xp}{x^{\prime}}
\newcommand{\xz}{x_0}
\newcommand{\xzp}{x_0^{\prime}}

\begin{document}
\begin{center}
\Large{HISTORIES APPROACH TO QUANTUM MECHANICS}

\vspace{.2in}
\large{Tulsi Dass}

\vspace{.12in} 
Chennai Mathematical Institute

92, G.N. Chetty Road, T. Nagar, Chennai, 600017, India 

Email : tulsi@cmi.ac.in
\end{center}

\vspace{.15in}
\noindent
Extended version of lectures in the Nineth Frontier Meeting at the Centre for 
Learning at Varadanahalli (near Bangalore), Jan 3-7,2005.

\vspace{.2in}
\noindent
\textbf{Contents}

\begin{enumerate}

\item  Introduction  
\item  The $\Pi$ Functions of Houtappel, Van Dam and Wigner; Primitive Concepts of 
Physical Theories 
\item  The Consistent History Formalism of Griffiths and Omnes \\
3.1 Formula for a temporal sequence of measurement results in standard 
quantum theory \\
3.2 The decoherence functional in traditional quantum mechanics \\
3.3 Griffiths' consistent history scheme \\
3.4 Refinement and compatibility; Complementarity \\
3.5 Example
\item  Understanding the Quasiclassical Domain \\
4.1 The programme \\
4.2 Path integral representation of the decoherence functional \\ 
4.3 Classical equations of motion in linear quantum systems 
\item  Generalized Histories-based Quantum Mechanics; Quantum Mechanics of 
Space-time \\ 
5.1 The scheme of generalized quantum mechanics \\ 
5.2 Quantum mechanics of space-time  
\item  Some  Developments in the Formalism of Histories-based Theories \\
6.1 Generalization of the notion of time sequences; partial semigroups \\
6.2 Quasitemporal structures \\ 
6.3 The history projection operator (HPO) formalism \\
6.4 The algebraic scheme of Isham and Linden \\
6.5 An axiomatic scheme for quasitemporal histories-based theories; 
Symmetries and conservation laws in histories-based theories

\end{enumerate}

\noindent
I. \textbf{Introduction}

\vspace{.12in}
Quantum mechanics (QM), in its traditional formulation, is an eminently 
successful theory. It has, however, some unsatisfactory features. The 
experimentally verifiable predictions of the theory (which are generally 
probabilistic) about any quantum system are made with reference to an external 
 observer who can test those predictions by repeatedly performing an experiment with prescribed initial conditions and invoking the frequency interpretation 
 of probability. Such a formalism is clearly inadequate for the treatment of 
 quantum dynamics of the universe as a whole. 

 Another unsatisfactory feature of the formalism is the postulate of state 
vector reduction in the theoretical treatment of quantum measurements. Its 
ad-hoc nature has always appeared unsatisfactory to those who care not only 
about the empirical adequacy of a theory but also about its intuitive appeal 
and conceptual coherence. A satisfactory formalism for QM should either 
provide an explanation of this reduction or else circumvent it.

The consistent history approach to QM is a reformulation/reinterpretation and 
extension  of the traditional formalism of QM which is applicable to closed 
systems (in particular, the universe) and avoids the postulate of 
state vector reduction. The central objects in this approach are `histories' 
which are temporal sequences of propositions about properties of a physical system. I shall try to cover the essential developments in this area 
allowing the choice of topics for the last one-third of the course to be 
somewhat subjective.

\vspace{.2in}
\noindent
II. \textbf{The $\Pi$ Functions of Houtappel, Van Dam and Wigner; Primitive 
Concepts of Physical Theories}

\vspace{.12in}
Objects analogous to what are now called histories were used way back in 
mid-sixties by Houtappel, Van Dam and Wigner (HVW) [1] who employed, in 
an article 
entitled `The conceptual basis and use of geometric invariance principles', 
conditional probabilities $\Pi(B|A)$ where \linebreak $A = (\alpha, r_{\alpha}; 
\beta, r_{\beta};...;\epsilon, r_{\epsilon})$ denotes a set of measurements 
$ \alpha, \beta,...,\epsilon $ (at times $t_{\alpha}, t_{\beta},...,
t_{\epsilon})$ with outcomes $r_{\alpha}, r_{\beta},...,r_{\epsilon}$ and 
similarly $B = (\zeta, r_{\zeta};...; \nu, r_{\nu}).$ The quantity $\Pi(B|A)$ 
denotes the probability of realization of the sequence B conditional to the 
realization of A. [I have changed HVW's notation $\Pi(A|B)$ to $\Pi(B|A)$ to 
bring it in line with the standard usage in probability theory.] The ordering 
of times $ t_{\alpha},...,t_{\epsilon}, t_{\zeta},...,t_{\nu}$ is arbitrary. 
The objects $\Pi(B|A)$ are quite general and can be employed in classical as 
well as quantum mechanics. The measurements refer to external observers.

 The central idea behind the HVW paper was `... to give a formulation of 
 invariances directly in terms of the primitive concepts of physical theory, 
 i.e. in terms of observations and their results.'A more systematic 
 identification of primitive elements of physical theories was done in ref [2]; 
 they are : \\
 (i) Observations/measurements \\
 (ii) Evolution of systems (this involves the concept of time--discrete or 
 continuous) \\
 (iii) Conditional predictions about systems: given some information about a 
 system, to make predictions/retrodictions about its behaviour. \\ 
 These three elements must be present (explicitly or implicitly) in every 
 physical theory --- classical, quantum or more general.

 \noindent
 Note. It is important to note here that, when basics are formalised in this 
 manner, the concept of \emph{information} enters quite naturally at a very 
 early  stage in the foundations of physics. 

 The concept of \emph{state} serves to integrate these three elements : a state is usually defined in terms of observable quantities; evolution of systems is 
 generally described in terms of change of state with time and, in the formal 
 statement of a problem of prediction/retrodiction, one generally gives 
 information about a system by specifying an initial state. 

 The $\Pi$ functions of HVW are objects defined in terms of primitive elements 
 of physical theory. The same is also true of histories, as we shall se below. 
 This, in the author's view, is the most important feature of history theories.

\vspace{.2in}
\noindent
\textbf{III. The Consistent History Formalism of Griffiths and Omnes [3-10]}

\vspace{.12in}
 \textbf{3.1 Formula for the probability of a temporal sequence of measurement 
results in standard quantum theory [1,11]}

\vspace{.12in}
We consider, for a typical quantum system S with initial state given by a 
density operator $\rho(t_0)$, a sequence of measurements at times $ t_1, 
t_2,...,t_n (t_0 < t_1 < t_2 ... <t_n)$. Let the measurement results at these 
times be represented by projection operators $P_1, P_2, ...,P_n$ (which may 
refer to, for example, a particle's position in a certain domain, its 
momentum 
in a certain domain, its spin projection in a certain  direction having a 
certain value, etc). We wish to calculate the joint probability 
\begin{eqnarray}
p = Prob(P_1,t_1;...;P_n,t_n|\rho(t_0))
\end{eqnarray}
of these measurement results. Let 
\begin{eqnarray}
\rho(t_0) = \sum_{i=1}^{k} w_i |\phi_i><\phi_i| \hspace{.2in} w_i \geq 0; 
\hspace{.15in} \sum w_i = 1.
\end{eqnarray}
We shall first calculate this probability with $\rho(t_0) = |\phi><\phi| $ 
(pure state) and take (2) into consideration at the end.

The two main ingredients in this calculations are : \\
(i) Between two consecutive measurements at times $t_j,t_{j+1}$, the system has unitary evolution given by the operator $U(t_{j+1},t_j)$. \\
(ii) At the completion of each measurement, we shall invoke the reduction 
postulate by applying the appropriate projection operator.

The (normalised) state just before the measurement at time $t_1$ is 
$ |\psi(t_1)> = U(t_1,t_0)|\phi>$. Probability of a measurement result 
corresponding to $ P_1 =1$ is given by \\
\begin{eqnarray}
p_1 = tr(P_1 |\psi(t_1)><\psi(t_1)|) = <\phi|U^{\dagger}(t_1,t_0)P_1U(t_1,t_0)
|\phi> = <\phi|P_1(t_1)|\phi>
\end{eqnarray}
where we have introduced the Heisenberg picture projection operators 
\begin{eqnarray}
P_j(t_j) = U^{\dagger}(t_j,t_0)P_jU(t_j,t_0) \hspace{.2in} j = 1,2,..,n.
\end{eqnarray}
Immediately after the measurement, the state is 
\begin{eqnarray}
D_1^{-1}P_1U(t_1,t_0)|\phi>
\end{eqnarray}
where 
\begin{eqnarray}
D_1 = <\phi|U^{\dagger}(t_1,t_0)P_1U(t_1,t_0)|\phi>^{1/2} = 
<\phi|P_1(t_1)|\phi>^{1/2} = p_1^{1/2}.
\end{eqnarray}
The state just before the measurement at time $t_2$ is 
\begin{eqnarray*}
|\psi(t_2)> = D_1^{-1}U(t_2,t_1)P_1U(t_1,t_0)|\phi>.
\end{eqnarray*}
The probability of a measurement result $ P_2 =1$ at time $t_2$ (conditional 
on the measurement result $P_1=1$ at time $t_1$ ) is 
\begin{eqnarray*}
p_2 & = & tr(P_2 |\psi(t_2)><\psi(t_2)|) \\
    & = & D_1^{-2}<\phi|U(t_1,t_0)^{\dagger}P_1U(t_2,t_1)^{\dagger}P_2U(t_2,t_1)          P_1U(t_1,t_0)|\phi> \\ 
    & = & D_1^{-2}<\phi|P_1(t_1)P_2(t_2)P_1(t_1)|\phi>.
\end{eqnarray*}
The joint probability of $P_1 = 1$ at time $t_1$ and $P_2 =1$ at time $t_2$ is 
\begin{eqnarray}
p_1p_2 & = &  <\phi|P_1(t_1)P_2(t_2)P_1(t_1)|\phi> \\ \nonumber
       & = & Tr(P_1(t_1)P_2(t_2)P_1(t_1)|\phi><\phi|) \\ \nonumber 
       & = & Tr(C_2|\phi><\phi|C_2^{\dagger})
\end{eqnarray}
where $ C_2 = P_2(t_2)P_1(t_1)$. Denoting the temporal sequence$ (P_1,t_1),..,
(P_n,t_n)$ by $\alpha$ and defining 
\begin{eqnarray}
C_{\alpha} = P_n(t_n)P_{n-1}(t_{n-1})...P_1(t_1)
\end{eqnarray}
we have, by a continuation of the argument above, the joint probability of the 
outcomes $\alpha$ with the initial state $|\phi>$ is (denoting by $p_j$ the 
probability of $P_j = 1$ at time $t_j$ conditional on $P_i = 1$ at times $t_i$ 
for i = 1,2,..,j-1)
\begin{eqnarray}
p_1p_2...p_n = Tr(C_{\alpha}|\phi><\phi|C_{\alpha}^{\dagger}).
\end{eqnarray}
Finally, the desired probability (1) is 
\begin{eqnarray}
p = Prob(\alpha|\rho(t_0)) = Tr(C_{\alpha}\rho(t_0)C_\alpha^{\dagger}).
\end{eqnarray}

\vspace{.15in}
\noindent
\textbf{3.2 The decoherence functinal in traditional quantum mechanics}

Let $\alpha$ be the temporal sequence $(P_1,t_1;P_2,t_2;...;P_n,t_n)$ of the 
previous subsection and $\beta$ the same as $\alpha$ except that, for a fixed 
$t_j$, $P_j$ is replaced by another projector $Q_j$ orthogonal to it ($P_jQ_j = 
Q_jP_j = 0$). Considering $\alpha$ and $\beta$ as propositions [ so that, for 
example, $Prob(\alpha|\rho(t_0))$ is the probability of the proposition 
$\alpha$ being true (with the initial state given to be $\rho(t_0)$)], let $\gamma$ be the proposition `$\alpha$ or $\beta$'; 
this is obtained by replacing $P_j$ in $\alpha$ by $P_j + Q_j$. Clearly 
\begin{eqnarray}
C_{\gamma} = C_{\alpha} + C_{\beta}.
\end{eqnarray}
Eq(10) gives
\begin{eqnarray}
Prob(\gamma|\rho(t_0)) & = & Tr(C_{\gamma}\rho(t_0)) \\ \nonumber
                       & = & Prob(\alpha|\rho(t_0)) + Prob(\beta|\rho(t_0)) + 
		       d(\alpha,\beta) + d(\beta,\alpha)
\end{eqnarray}
where 
\be
d(\alpha,\beta) = Tr(C_{\alpha}\rho(t_0)C_{\beta}^{\dagger}).
\ee
Note that
\be
d(\alpha,\beta) = d(\beta,\alpha)^{*} \\
d(\alpha,\alpha) \geq 0 
\ee
and 
\be
Prob(\alpha|\rho(t_0)) = d(\alpha,\alpha).
\ee

If $ \alpha$ and $\beta$ are to be mutually exclusive events in the sense of 
probability theory, we must have 
\be
Prob(\gamma|\rho(t_0)) =Prob(\alpha|\rho(t_0)) + prob(\beta|\rho(t_0)).
\ee
The last two terms in eq(12) which prevent eq(17) from holding good, represent 
quantum interference effects. Eq(17) holds if and only if the following 
`decoherence condition' is satisfied :
\be
Re d(\alpha,\beta) = 0.
\ee
The object $d(\alpha,\beta)$ is referred to as the \emph{decoherence 
functional}. We will have much to do with this object later. Apart from the 
important equations (16) and (18), a significant feature of this object is 
that it contains information about the Hamiltonian H and the initial state 
$\rho(t_0)$. [If it is desired to make this explicit, we can replace the 
term $d(\alpha,\beta)$ by $d_{H,\rho(t_0)}(\alpha,\beta)$.] This fact 
becomes of considerable importance in quantum cosmology where, in the context 
of quantum mechanics of the universe, the decoherence functional has 
information about the dynamics and the initial state of the universe.
                      
\noindent
Remark. Note that, in the traditional formalism of  quantum mechanics, 
probabilities of histories can be calculated without bothering about 
decoherence conditions like (18). (This will not be true in the Griffiths' 
consistent history scheme presented in the next subsection.) These conditions 
are needed if we wish to construct a probability space of histories. The 
situation here is somewhat similar to the calculation of the probability of a 
sequence of coin-tossing results which can be done without formally defining a 
probability space whose elements are coin-tossing sequences.
		      
\vspace{.2in}
\noindent
\textbf{3.3 Griffiths' consistent history scheme}

\vspace{.12in}
In Griffiths' scheme, one retains the kinematical Hilbert space framework of 
traditional quantum mechanics and the quantum dynamical equations of 
Heisenberg and Schr$\ddot{o}$dinger; the collapse postulate is discarded. It 
operates with closed quantum systems; measurement situations ar accomodated 
by considering the system plus apparatus as a closed system. Quantum mechanics 
is treated as a statistical theory which assigns probabilities to families of 
temporal sequences of the form $\alpha,\beta$ considered above (but dropping the condition that actual measurements are performed at times $t_1, t_2,..,t_n$); 
they are referred to as \emph{histories}.

The developments below refer to a closed quantum mechanical system S which 
has associated with it a Hilbert space $\mathcal{H}$. All operators and states 
are in the framework of this Hilbert space.

The basic interpretive unit is a \emph{quantum event} consisting of a pair 
(t,P) where t is a point of time and P a projection operator describing a 
possible state of affairs (also referred to a quantum property) at time t. 
The negation of (the proposition corresponding to) the event e = (t,P) (i.e. 
`not e') is the event $ e^{\prime}$ = (t, I-P). At any time t, one can define 
families of mutually exclusive and exhaustive sets of quantum events 
represented by mutually orthogonal projection operators constituting 
decompositions of unity :
\be
P^{(a)}_jP^{(b)}_j = \delta_{ab}P^{(a)}_j, \hspace{.2in} \sum_{a}P^{(a)}_j = I
\ee
where the index j labels a family. A family typically corresponds to a definite observable (or, more generally, 
to a set of mutually commuting observables), different members of a family 
being projection operators corresponding to non-overlapping domains in the 
(joint) spectrum of the observable(s), covering together the whole of the 
spectrum. 

As indicated above, a \emph{history} of the system S consists of a prescribed 
initial state $\rho(t_0)$ at some time $t_0$ and a time-ordered sequence of 
quantum events $(t_1,P_1),...,(t_n,P_n)$ where 
\be
t_0 < t_1 < ... < t_n.
\ee

\noindent
Note. Quite often, the initial state is understood (or pescribed separately) 
and history is defined as a time-ordered sequence of quantum events. This is 
convenient in, for example, quantum cosmology where the initial state of the 
universe is supposed to be fixed. We shall allow ourselves to be flexible in 
this matter.

To assign a probability to a history $\alpha$, we must consider an exhaustive 
set of mutually exclusive histories (containing $\alpha$) such that 
probabilities assigned to the histories in the set add up to unity. The 
standard way of constructing such a family is to fix a time sequence [say, the 
one in eq(20)] and the initial state $\rho(t_0)$ and construct, for each $t_j$, 
a decomposition of unity $P_j^{(\alpha_j)}$ satsfying eqs(19) 
[with $a = \alpha_j$ 
etc]; the desired family consists of all possible sequences
\be
(t_1,P_1^{(\alpha_1)}),(t_2,P_2^{(\alpha_2)}),...,(t_n,P_n^{(\alpha_n)}).
\ee
Note that there is a freedom of choosing different partitions of unity at 
different times. The history represented by the sequence (21) is labelled by 
the symbol $\alpha = (\alpha_1,...,\alpha_n)$ which serves to distinguish 
different members of the family; the history itself is often referred to as 
$\alpha$ or by another symbol like $h_{\alpha}$.
 
Some applications require more general histories ( the so-called 
`branch-dependent' histories) in which the choice of the projector 
 at time $t_j$ depends on the projectors at earlier times 
$t_1, t_2, ...,t_{j-1}$. Such a history may be represented as a sequence 
of the form
\be
(t_1,P_1^{(\alpha_1)}),(t_2,P_{(2,\alpha_1)}^{(\alpha_2)}),...,
(t_n,P_{(n,\alpha_1,..,\alpha_{n-1})}^{(\alpha_n)}).
\ee
Unless stated otherwise, we shall work with histories of the form (21). We 
shall refer to the time sequence $(t_1,...,t_n)$ as the \emph{temporal 
support} (or simply \emph{support}) of the history $\alpha$.

Coarse-graining of histories is often employed in theoretical work. There are 
several ways of doing this. We mention here the simplest coarse-graining in 
which the temporal supports are not changed (other methods will be described 
when  needed); this is done by replacing one or more projectors 
$P_j^{(\alpha_j)}$ by the `coarser' projectors 
\be
Q_j^{(\beta_j)} = \sum_{\alpha_j}b_{\alpha_j}^{\beta_j}P_j^{(\alpha_j)} \hspace{.2in} 
(b_{\alpha_j}^{\beta_j} = \textnormal{0 or 1})
\ee

Defining the Heisenberg projectors as in eq(4) (the time $t_0$ in that 
equation can, in fact, be chosen as some arbitrary reference time $t_r$), we 
define the \emph{chain operator} for a history $\alpha$ [see eq(8)] as
\be
C_{\alpha} = P_n^{(\alpha_n)}(t_n)...P_2^{(\alpha_2)}(t_2)P_1^{(\alpha_1)}
(t_1).
\ee
Relations (19) (with appropriate replacement of indices) imply
\be
\sum_{\alpha}C_{\alpha} = I
\ee
which represents the completeness of the family $\mathcal{F} = \{\alpha\}$.

Given two histories $\alpha$ and $\beta$ in the family $\mathcal{F}$, the 
decoherence functional $d(\alpha,\beta)$ is defined as in eq(13) [with 
$C_{\alpha}$ given by eq(24)]; it satisfies the conditions (14) and (15). It 
is a measure of the quantum interference between the two histories $\alpha$ 
and $\beta$.(This should be clear fom section 3.2.) The two histories $\alpha$ 
and $\beta$ are said to satisfy the \emph{consistency condition} (or 
\emph{decoherence condition}) if eq(18) holds. When this condition holds for 
every pair of histories in the family $\mathcal{F}$, it is called a 
\emph{consistent family}. [The word \emph{framework} has been coined by 
Griffiths[12] for a consistent family of histories.] Legitimate probability 
assignments can be made only 
for histories in a consistent family. 

Given a consistent family $\mathcal{F}$, the probabilty of a history $\alpha$ 
in $\mathcal{F}$ is \emph{postulated} to be given by eq(16). Equations (25), 
(14) and (18) ensure that 
\be
\sum_{\alpha}Prob(\alpha) = \sum_{\alpha}d(\alpha,\alpha) = 1.
\ee

It was found in most applications that, whenever the decoherence condition 
(18) holds, the stronger condition
\be
\dab = 0 , \hspace{.2in} \alpha \neq \beta 
\ee
also holds. Being convenient to apply, it is the condition (27) that is more 
often used. Eq(18) is referred to as the \emph{weak decoherence condition} 
and (27) as the \emph{medium strong decoherence condition}. For the 
treatment of the so-called `strong decoherence condition', the reader is 
referred to ref[13].

It must be noted, however, that, whereas the conditions like (27) or 
stronger ones are sufficient to ensure consistency, eq(18) is a necessary 
and sufficient condition for it. The latter must be used if some question of 
principle or finer point is to be settled.

\noindent
Note. Adoption of the formula (16) (which was \emph{derived} in the traditional 
formalism of quantum mechanics \emph{by employing the reduction postulate} 
as a \emph{postulate} in the history theory might make a critical reader 
suspicious.``What is the great point of doing all this if the reduction 
postulate is being allowed to enter `from the back door' in the formalism 
like this ?'' she/he might ask. The point, stated in one sentence, is that 
the procedure being adopted constitutes a valid probability assignment 
(and, relatedly, a legitimate realization/representation of the relevant 
Boolean lattice/logic) -- no 
matter in what context the formula being adopted first appeared.

Let me explain/elaborate.

In probability theory, one has a probability space consisting of a triple 
$(\Omega,\mathcal{B},P)$ where $\Omega$ is a nonempty set (sample space or the 
space of elementary events), $\mathcal{B}$ (the space of events) a family of 
subsets of $\Omega$ closed under unions, intersections and complements and 
P  a real-valued function on $\mathcal{B}$ (assignment of probability to 
events) satisfying the conditions \\
(i) $P(A) \geq 0$ for all $A \in \mathcal{B}$ \\
(ii)$ P(\emptyset) = 0$ and $ P(\Omega) = 1$ \\
(iii) $P(A \cup B) = P(A) + P(B)$ if $ A \cap B = \emptyset$.   
When elements of $\mathcal{B}$ are interpreted as propositions (for example, 
in the context of coin tossing, the event `head', when expressed as `The head 
appears on tossing the coin' is a proposition), the operations of union, 
intersection and complementation in $\mathcal{B}$ correspond, respectively, to 
disjunction ($A \cup B$ corresponds to `A or B' usually written $A \vee B$), 
conjunction ($A \cap B $ corresponds to `A and B' usually written 
$A \wedge B$) and negation ($A^c$ corresponds to `not A' generally written as 
$\neg A$ or $A^*$); one then refers to $\mathcal{B}$ as a Boolean logic or a 
Boolean lattice.

All probabilistic reasoning (whether applied to systems on which repeatable 
experiments can be performed or to closed systems like the universe) must be 
based on bonafide representations/realizations of Boolean logic. (In logic, 
such a realization is referred to a `universe of discourse'.) In the usual 
applications  of probability theory, this requirement is (generally 
implicitly) taken care of.

The justification of the above-mentioned assignment of probabilities to 
consistent histories is that a consistent family constitutes a bonafide 
representation of Boolean logic. For details, see Omnes [4,6,9]. 

\vspace{.15in}
\noindent
\textbf{3.4 Refinement and compatibility; Complementarity}

\vspace{.12in}
We shall use the letters $\mathcal{F}, \mathcal{G},...$ to denote consistent 
history families (CHFs). A CHF $\mathcal{G}$ is said to be a 
\emph{refinement} of another CHF $\mathcal{F}$ (or $\mathcal{F}$ is a 
\emph{coarsening} of 
$\mathcal{G}$) if \\
(i) supp($\mathcal{G}$) contains supp($\mathcal{F}$) (i.e. the time-points on 
which $\mathcal{G}$ is defined include those on which $\mathcal{F}$ is 
defined); \\
(ii) at each time in supp($\mathcal{F}$), the decomposition of unity employed 
in $\mathcal{G}$ is either the same as that employed in $\mathcal{F}$ or 
finer. 

Two CHFs $\mathcal{F}$ and $\mathcal{F}^{\prime}$ are said to be 
\emph{compatible} if there is a CHF $\mathcal{G}$ which is a refinement of 
both $\mathcal{F}$ and $\mathcal{F}^{\prime}$; otherwise they are said to be 
\emph{incompatible}.

More than one CHFs can describe the same physical process without being 
mutually compatible. (We shall see an example below.) This phenomenon is called \emph{complementarity} and the mutually incompatible histories involved in 
the description are said to be complementary. This is a concrete formulation 
of Bohr's principle of complementarity in the consistent history framework.
Its origin lies, of course, in the noncommutativity of the operator algebra 
in quantum kinematics (and the consequent possibility of constructing 
mutualy noncommuting  complete sets of commuting observables). For more 
details, see [9].

\vspace{.15in}
\noindent
\textbf{3.5 Example[9]}

We consider a free spin 1/2 particle with the Hamiltonian H = 0 (which means 
that the 
translational degrees of freedom are suppressed). Let $ t_0 < t_1 < t_2. $ 
The initial state at time $t_0$ is the $\vec{S}.\hat{n}_0 = 1/2 $ state 
corresponding to $ \rho(t_0) = \frac{1}{2}(I + \vec{\sigma}.\hat{n}_0)$. At 
time $t_2$, the spin component $\vec{S}.\hat{n}$ is measured with the result 
+1/2. We wish to study as to what can be said about $\vec{S}.\hat{n^{\prime}}$ 
at an intermediate time $t_1$.

We employ the following decompositions of identity at the times $t_1$ and 
$t_2$: \\
\begin{eqnarray*}
t_1 : I = P_{+}^{\prime} + P_{-}^{\prime}; \hspace{.15in} P_{\pm}^{\prime} 
= \frac{1}{2}(I \pm \vec{\sigma}.\hat{n^{\prime}}) \\
t_2 : I = P_{+} + P_{-}; \hspace{.15in} P_{\pm} = \frac{1}{2}(I \pm 
\vec{\sigma}.\hat{n})
\end{eqnarray*}

We consider two histories $\alpha$ and $\beta$ given by 
\begin{eqnarray*}
\alpha : (t_1, P_{+}^{\prime}), (t_2, P_{+}) \\
\beta : (t_1, P_{-}^{\prime}), (t_2, P_{+}).
\end{eqnarray*}
We have (recalling tha H = 0)
\begin{eqnarray*}
C_{\alpha} = P_{+}(t_2)P_{+}^{\prime}(t_1) = P_{+}P_{+}^{\prime} \\
C_{\beta} = P_{+}(t_2)P_{-}^{\prime}(t_1) = P_{+}P_{-}^{\prime}.
\end{eqnarray*}
The consistency condition (18) gives 
\begin{eqnarray*}
Re [ Tr (C_{\alpha}\rho(t_0)C_{\beta}^{\dagger})] = 0 
\end{eqnarray*}
which, after a straightforward calculation, gives the condition
\be
(\hat{n} \times \hat{n^{\prime}}).(\hat{n}_0 \times \hat{n^{\prime}}) = 0 
\ee
which is satisfied by $ \hat{n^{\prime}} = \pm\hat{n}_0, \pm \hat{n}$ and  
some other values. 

It is of interest to consider a couple of special cases :

\noindent
(i)$ \hat{n} = \hat{n}_0$. Eq(28) now gives $ \hat{n^{\prime}} \times \hat{n}_0 = 0 $ 
implying $ \hat{n^{\prime}} = \pm \hat{n}_0.$ In this case, probabilistic 
reasoning can give information only about the spin component 
$\vec{S}.\hat{n}_0$ at time $t_1$. Eq(16) gives (with $\hat{n^{\prime}} = 
\hat{n}_0 $ ) $Prob(\alpha) = 1$  and $Prob(\beta) = 0$ which is consistent 
with the prediction of the traditional formulation of  quantum mechanics. 
(In an eigenstate of an observable A with eigenvalue $\lambda$, a measurement 
of A always gives the value $\lambda$,)

\noindent
(ii) $ \hat{n}_0 = \hat{i}, \hat{n} = \hat{k}$. Eq (28) now gives 
$n_x^{\prime}n_z^{\prime} = 0$. This condition permits, among others, the 
following two families which are mutually incompatible (and, therefore, 
complementary): 
\begin{eqnarray*}
(a) (t_1, S_x = \pm 1/2), (t_2, S_z =1/2) \\
(b) (t_1, S_z = \pm 1/2), (t_2, S_z = 1/2).
\end{eqnarray*}

\vspace{.2in}
\noindent
\textbf{IV. Understanding the Quasiclassical Domain}

\vspace{.15in}
\noindent
\textbf{4.1 The programme [9,14]}

\vspace{.12in}
An important agenda item for any scheme of quantum mechanics of the universe 
is to explain the existence of the quasiclassical domain of everyday 
experience as an emergent feature of the quantum mechanical formalism. The 
term `quasiclassical' refers to the fact that the classical deterministic 
laws give only an approximate description of this domain. (For example, 
we need a nonzero Planck constant to develop the foundations of classical 
statistical mechanics properly.) We would like to understand how, in the 
course of the quantum evolution of the universe, certain subsystems-- a 
large majority of the macroscopic systems-- manage to have a 
near-deterministic evolutution and to have a clearer picture (than the 
deterministic classical laws provide) of this near-deterministic evolution.

In this section we shall  present the histories approach to this problem 
keeping close to the paper [15] of Gell-Mann and Hartle who build up and 
improve upon the works of some earlier authors [16,17,5,6,9].

The arena of classical mechanics of a subsystem S of the universe is the phase 
space $\Gamma$ of S. With a nonzero Planck contant, the uncertainty principle 
is operative which prohibits the  classical evolution of S in 
terms of phase space trajectories. Moreover, quantum mechanics is 
probabilistic whereas classical mechanics is deterministic. In the 
 consistent histories approach to  quantum dynamics of the universe, 
 what one hopes to show is that some appropriately chosen histories have, 
 with probability very near unity, \\
 (i) their `events' at various times  described by cells of near-
 minimal size in $\Gamma$ and \\
 (ii) the sequential locations of these cells are correlated by the classical 
 laws of evolution.

The description of quantum mechanics of the universe at the deepest level is 
expected to include interacting quantum fields, strings or more fundamental 
objects and a specification of the initial state of the universe. (Recall that 
the definition of decoherence functional in the quantum mechanics of a closed 
system involves, besides a choice of time points, decompositions of unity 
at the 
chosen time points and the law of evolution, the specification of an 
initial state.) The existence of the quasiclassical domain depends crucially 
on the fundamental interactions operative at deeper levels as well as the 
initial state of the universe. If we had a reasonably complete theory of the 
quantum dynamics of the universe, that theory would dictate a natural choice 
of various ingredients  for the histories to be chosen for the description of 
the quasiclassical domain. In the absence of such a theory, we shall be guided 
 mainly by the intuition based on experience with classical physics.
 
We shall assume a fixed space-time background ignoring the problems relating to the description of quantum dynamics of the geometry of space-time. We shall 
also assume a fixed time variable t. (This only means that we are operating in 
a fixed reference frame -- Galilean or Lorentzian-- and does not 
necessarily mean commitment to nonrelativistic mechanics.) The choice of time 
points in the histories to be considered will, therfore, be the usual one 
adopted in the previous section.

To decide on the decompositions of unity to be employed at the various time 
points, we must choose a (not necessarily complete) set of commuting 
observables. Typical classical observables of interest are pointer position 
on a measuring device, centre of mass of a cricket ball, etc. These are 
examples of the so-called \emph{collective observables}[5,6,9]. Classically, 
they are described in terms of  phase space variables. The  events of 
interest  (for the construction of the relevant histories) are represented 
classically by domains in the appropriate phase space which are neither too 
small nor too large. Construction of the corresponding quantum mechanical 
projection operators 
 (or more genarally, the so-called `quasi-projectors') is a nontrivial 
task. One approach to constructing such operators is to employ integrals of 
coherent state projectors $ |f_{qp}><f_{qp}|$ over appropriate domains D :
\be
F = (2\pi \hbar)^{-1}\int_D |f_{qp}><f_{qp}|dqdp.
\ee
For a detailed treatment of these matters, we refer to Omnes[5,6,9].

We shall model reality by introducing a cofiguration space for the closed 
universe cooordinatized by the configuration variables $q^{\alpha}$; these 
will be divided into the relevant variable $x^a$ and another set $Q^A$ 
which parametrise the configuration space of the `rest of the universe' 
(which acts essentially as a heat bath) and are to be ignored. The systems 
with configuration variables x and Q will be referred to as A and B 
respectively. The histories 
considered will employ decompositions of unity corresponding to 
non-overlapping  
exhaustive sets of domains of the configuration space parametrised by the 
variables $x^a$.

Interaction between the distinguished and the ignored variables causes 
decoherence of the  histories (chosen as above) by the rapid dispersal 
of the quantum mechanical phase information among the ignored variables 
[18,19]. (Details of this `environmental decoherence' will be taken up 
elsewhere.) These interactions, apart from producing the expected classical 
behaviour (of the system with configuration variables $x^a$ ) characterized 
by predictability, also produce noise. Suppression of the effect of this 
noise to achieve classical predictability is achieved by 
additional coarse-graining of histories [15].

We shall not consider situations involving chaos in nonlinear classical 
systems; for a treatment of this, see ref[20].

\vspace{.15in}
\noindent
\textbf{4.2 Path integral representation of the decoherence \\
functional[15,16,21,17,22]}

\vspace{.12in}
We shall now obtain a path integral representation for the decoherence 
functional $d(\alpha,\beta)$ of eq(13) where 
\begin{eqnarray*}
C_{\alpha} = P_n^{(\alpha_n)}(t_n)...P_1^{(\alpha_1)}(t_1) 
\end{eqnarray*}
and similarly for $C_{\beta}$. The projection operators $P_j^{(\alpha_j)}$
and $P_j^{(\beta_j)}$ represent the restriction of the configuration 
variables $q^{\gamma}$ to the domains $\Delta_j^{(\alpha_j)}$ and 
$\Delta_j^{(\beta_j)}$ respectively at the indicated times.

Suppressing the superscript of the configuration variables, let $q_S$ be the 
coordinate position operator in the Schr$\ddot{o}$dinger picture and 
\begin{eqnarray*}
q_H(t) = e^{iH(t-t_0)}q_Se^{-iH(t-t_0)} 
\end{eqnarray*}
the same in the Heisenberg picture. 
Their (generalized) eigenvectors satisfy, in obvious notation, the relations

\begin{eqnarray*}
q_S|\qp> = \qp |\qp> \hspace{.2in} q_H(t)|\qp,t> = \qp|\qp,t> \\
|\qp,t> = e^{iH(t-t_0)}|\qp> \\
\int d\qp |\qp><\qp| = I = \int d\qp |\qp,t><\qp,t|.
\end{eqnarray*}
The path integral repsentation of the Schr$\ddot{o}$dinger kernel is given by 
\begin{eqnarray*}
<\qdp , \tdp | \qp , \tp > = < \qdp |e^{-iH(\tdp - \tp )/\hbar} |\qp > 
= \int_{\tp , \qp}^{\tdp , \qdp}Dq e^{iS[q]/\hbar}
\end{eqnarray*}
where the integral Dq is over (an appropriate class of) configuration space 
trajectories q(t) satisfying the end-point conditions $q(t^{\prime}) = 
q^{\prime}$ and $ q(t^{\prime \prime} = q^{\prime \prime}.$ We shall write 
simply $\rho$ for $\rho(t_0)$.

Putting $t_n = T$, we have 

\noindent
$\dab  =  Tr(\ca \rho \cbd)$
\begin{eqnarray}  
=  \int \int \int \int dq_f^{\prime}dq_fdq_0^{\prime}
dq_0 \delta(q_f - q_f^{\prime})<q_f^{\prime},T|\ca|q_0^{\prime},0>
<q_0^{\prime},0|\rho|q,0><q_0,0|\cbd|q_f,T>. 
\end{eqnarray}
Now (taking, temporarily, n = 2)

\noindent
$<q_f^{\prime},T|\ca|q_0^{\prime},0>$
\begin{eqnarray}
& = & \int \int dq_1 dq_2 <q_f^{\prime}|P_2^{(\alpha_2)}|q_2><q_2|
      e^{-iH(t_2-t_1)/\hbar}|q_1>
      <q_1|P_1^{(\alpha_1)}|q_0^{\prime}>  \nonumber \\ 
& = & \int_{[q_0^{\prime}\alpha q_f^{\prime}]} Dq^{\prime}
     e^{iS[q^{\prime}]/\hbar}
\end{eqnarray}
where the notation in the subscript in the last line means that the 
integration is over 
configuration space trajectoties from $q_0^{\prime}$ to $q_f^{\prime}$ 
consistent 
with the constraints of the history $\alpha$ at its temporal support. Now we 
can remove the restriction n = 2; eq(31) continues to be valid. 

Proceeding similarly for the $\cbd$ term in eq(30), we have, finally, 

\noindent
$\dab   =    \int \int \int \int dq_f^{\prime} dq_f dq_0 dq_0^{\prime} 
             \delta (q_f - q_f^{\prime})  
             \int_{[q_0^{\prime} \alpha q_f^{\prime}]} 
              Dq^{\prime} \int_{[q_0 \beta q_f]} D \qdp 
	      e^{i(S[\qp] - S[\qdp])/\hbar} 
             \rho(q_0^{\prime} , q_0) $ 
\begin{eqnarray}	     
\equiv \int_{(\alpha)} D \qp \int_{(\beta)}D \qdp 
              e^{i(S[\qp] - S[\qdp])/\hbar}
              \rho(q_0^{\prime}, q_0).
\end{eqnarray}

Introducing the systems A and B as in the previous subsection and defining the 
reduced  density operator for the system of interest A as $\tilde{\rho}_A = 
Tr_{B} \rho$, we have, in the coordinate representation
\be
\tilde{\rho}_A(x_0^{\prime}, x_0) = \int dQ_0 \rho(x_0^{\prime}, Q_0; x_0, Q_0).
\ee
Putting 
\be
S[q] = S_A[x] + S_B[Q] + S_I[x,Q]
\ee
and $D \qp = Dx^{\prime} DQ^{\prime}$ etc in eq(32), we carry out the 
Q-integrations. Defining the \emph{influence phase} W by the relation 

\noindent
$ e^{iW[x^{\prime},x;x_0^{\prime},x_0)} 
\tilde{\rho}_A (x_0^{\prime},x_0) $ 
\begin{eqnarray}
\equiv \int \int DQ^{\prime}DQ \delta(Q_f^{\prime} - Q_f)
exp \{\frac{i}{\hbar}(S_B[Q^{\prime}] + S_I [x^{\prime},Q^{\prime}]
-S_B[Q] - S_I[x,Q]) \} \\ \nonumber 
\rho(x_0^{\prime},Q_0^{\prime};x_0,Q_0)
\end{eqnarray}
(where the notation W[.,.;.,.) means that it is a functional of the first two 
arguments and an ordinary function of the last two) we have 
\be
\dab = \int_{\alpha} \int_{\beta} D \xp Dx \delta(\xp_f - x_f)
e^{i(S_A[\xp] - S_A[x] + W[\xp,x;\xzp,\xz))/\hbar} \tilde{\rho}_A (\xzp, \xz).
\ee
The influence phase W (which is a slight generalization of the quantity with 
the same name in ref[16]) represents the influence of the heat bath B on the 
dynamics of the sytem A  in terms of the variables of A. 

Generalizations of eqs(32) and (36) for the case when  the histories $\alpha$ 
and $\beta$ involve restrictions to domains of generalized momenta as well 
may be found in ref[15].

\vspace{.15in}
\noindent
\textbf{4.3 Classical equations of motion in linear quantum systems}

\vspace{.12in}
We shall consider the model of eq(34) supplemented with the following 
additional assumptions : \\
(i)The action $S_A$ is of the form 
\be
S_A[x] = \int_0^T dt [\frac{1}{2}\dot{x}^T(t)M\dot{x}(t) - 
\frac{1}{2}x^T(t)Kx(t)]
\ee
where the superscript T denotes matrix transpose. \\
(ii) The part $S_I$ is local in time and linear in x : 
\be
S_I[x,Q] = \int_0^T dt x^T(t)f(Q(t)) 
\ee
where f(Q) is linear homogeneous in the Qs. \\
(iii) The initial density matrix factorizes :
\be
\rho(\xzp,Q_0^{\prime}; \xz,Q_0) = \rho_A(\xzp,\xz) \rho_B(Q_0^{\prime}, Q_0).
\ee
Eqs(33) and (39) give 
\be
\tilde{\rho}_A(\xzp,\xz) = \rho_A(\xzp,\xz).
\ee
(iv) The initial density matrix of the system B is of the form 
\be
\rho_B(Q_0^{\prime},Q_0) = e^{-B(Q_0^{\prime},Q_0)}
\ee
where B(.,.) is a bilinear form. 

Under these conditions, W has no explicit dependence on $\xzp$ and $\xz$ and 
has the  general form 

\noindent
$ W[\xp, x]
 =  \frac{1}{2}\int_0^T dt \int_0^t d\tp [\xp(t)-x(t)]^T 
      [k(t, \tp) \xp(\tp) +k^*(t, \tp) x(\tp)] $  
\begin{eqnarray} 
 =  \frac{1}{2} \int_0^T dt \int_0^t d\tp [\xp(t) - x(t)]^T  
      \{ k_R(t,\tp)[\xp(\tp)+x(t)] + ik_I(t,\tp)[\xp(\tp) - x(t)] \}.
\end{eqnarray}

\noindent
Here $ k(.,.) = k_R(.,.) + ik_I(.,.) $ is a complex matrix kernel. Moreover [23] 
\be
Im W[x,\xp] \geq 0.
\ee

Putting
\be
X(t) = \frac{1}{2}[\xp (t) + x(t)] ; \hspace{.2in} \xi(t) = \xp (t) - x(t)
\ee
we have, from eq(36), 
\be
\dab = \int_{\alpha} D \xp \int_{\beta} Dx \, d_1[X,\xi] 
\ee
with
\be
d_1[X, \xi] = \delta(\xi_f) e^{iA[X, \xi ]/\hbar} 
\rho_A(X_0 + \frac{\xi_0}{2}, X_0 - \frac{\xi_0}{2})
\ee
where
\be 
A[X, \xi] = S_A[X+ \frac{\xi}{2}] - S_A[X - \frac{\xi}{2}] + W[X, \xi]. 
\ee
After a few integrations by parts, we have 
\be
A[X, \xi] = -\xi^T_0M\dot{X_0} + \int_0^T dt \xi^T(t) e(t,X] + 
\frac{i}{4}\int_0^T dt \int_0^T d\tp \xi^T(t) k_I(t, \tp) \xi(\tp)
\ee 
where
\be
e(t,X] = -M\ddot{X}(t)  - KX(t) + \int_0^t d \tp k_R (t, \tp)X(\tp).
\ee

Eq(43) implies that $d_1$ has a decreasing exponential $exp[-Im W/\hbar]$. 
It can be shown [16] that, with appropriate choice of domains 
$\Delta_j^{(\alpha_j)}$ and $\Delta_j^{(\beta_j)}$, significant contributions 
to $\dab$ in eq(45) can come only come from $\xi(t)$ near zero and for 
$\alpha = \beta$. This implies, to a good approximation, the medium 
decoherence condition 
\be
\dab \approx 0, \hspace{.15in} \alpha \neq \beta. 
\ee
We can now define the probability of a history in the decoherent family :
\be
p(\alpha)& = & d(\alpha, \alpha) \\ \nonumber
& \cong& \int_{\alpha} DX [det(k_I/4\pi)]^{-1/2}. \\ \nonumber
&.& exp [ -\frac{1}{\hbar} \int_0^Tdt \int_0^T d\tp e^T(t, X] 
(k_I)^{-1} (t, \tp) e(\tp, X] ] w_A(X_0, M\dot{X}_0). 
\ee
Here $ w_A(X, M \dot{X})$ is the Fourier transform of 
$ \rho_{A}(X_0 + \frac{\xi_0}{2}, X_0 + \frac{\xi_0}{2})$ with respect to 
$\xi_0$ (Wigner function). 

It is clear from eq(51) that the histories with the largest probabilities 
will be those for which 
\be
0 =e(t,X] = -M \ddot{X}(t) - K X(t) + \int d \tp k_R (t, \tp) X(\tp).
\ee
This equation differs from the relevant Euler-Lagrange equation from the 
assumed action by the last term which is nonlocal in time; it arises from 
the interaction of the system A with the bath B. 

When the bath B is treated as a collection of simple harmonic oscillators 
in thermal equilibrium at some temperature $T_B$, 
explicit expressions for the influence phase can be obtained. Replacing 
discrete bath oscillators by  a continuum of oscillators with a cutoff 
frequency $\Omega$ and going to th Fokker-Planck limit [17] 
\be
kT_B \gg \hbar \Omega \gg \hbar \omega_R 
\ee
(where $\omega_R$ is the frequency of the distinguished oscillator 
renormalized by its interaction with the bath), eq(52) takes the familiar form 
\be
e(t,X] \simeq - M \ddot{X}(t) -K X(t) - 2M \gamma \dot{X}(t) = 0.
\ee

In eq(51), individual classical histories are distributed according to the 
probabilitis of their initial conditions given by the Wigner function 
$w_A(X_0, M \dot{X_0})$. The fact that the Wigner function is not, in 
general, non-negative need not cause concern because positive definiteness 
of the probability $p(\alpha)$ has already been guaranteed earlier. 

Eq(51) also reflects deviations from classical predictability. These 
deviations arise due to noise -- the source of which is the same as the one responsible for decoherence -- interaction with the bath B. For detailed treatment 
of this and related features, see[15].

\vspace{.2in}
\noindent
\textbf{V. Generalized Histories-based Quantum Mechanics ; Quantum 
Mechanics of Space-time[14,24,25]}

\vspace{.15in}
\noindent
\textbf{5.1 The scheme of generalized quantum mechanics} 

\vspace{.12in}

Application of the concept of history as a time-sequence of projection 
operators to quantum cosmology would involve facing the problem of time 
which arises due to non-availability of a preferred family of space-like 
hypersurfaces in quantum gravity. To bypass this problem, 
Hartle introduced a generalization of the history version of quantum 
mechanics in which histories are taken as the fundamental entities (without 
any reference to time sequences or projectin operators) and decoherence 
functionals are introduced by taking their properties noted earlier as 
defining properties. 
Formally, the generalized quantum mechanics of a closed system is defined by 
the following three ingrdients : 

\noindent
(1) Fine-grained histories :  These are objects representing the most refined 
description of dynamical evolution of the closed system to which one can 
contemplate assigning probabilities. For probability assignments, one 
considers sets \{f \}of fine-grained histories which are exclusive and 
exhaustive. Typical examples of fine-grained histories are the set of 
(continuous) particle paths in nonrelativistic quantum mechanics and the 
set of field configurations in space-time in quantum field theory. 

\noindent
(2) Allowed coarse-grainings : A concrete scheme of generalized quantum 
mechanics generally permits a restricted class of possible coarse-grainings 
of histories. For example, the generalized quantum mechanics of gauge theories 
will permit only gauge-invariant classes of histories as coarse-grained 
histories. 

\noindent
(3) Decoherent functionals : A decoherence functional is  a complex-valued 
function defined on pairs of histories in an exhaustive coarse-grained set 
$ \{ \alpha \}$ (which, as a special case, may be the set \{ f \} of 
fine-grained histories) satisfying the following conditions [ see 
eqs(14, 15,26)] \\
(i) Hermiticity : 
\be
\dab^* = d (\beta, \alpha);
\ee
(ii) positivity : 
\be
d(\alpha, \alpha) \geq 0;
\ee
(iii) normalization :
\be
\sum_{\alpha, \beta} \dab = 1; 
\ee
(iv) biadditivity : Given an (exclusive, exhaustive) family 
$ \{ \bar{\alpha} \} $ which is obtained by (further) coarse-grainingof a 
(possibly coarse-grained)family $ \{ \alpha \} $, we have 
\be
d(\bar{\alpha}, \bar{\beta}) = \sum_{\alpha \in \bar{\alpha}} 
\sum _{\beta \in \bar{\beta}} \dab.
\ee

Probabilities are assigned only to histories in a family $ \{ \alpha \} $ 
satisfying (generally approximately) a decoherence condition (with respect 
to some fixed decoherent functional d). Typical decoherence conditions are 
the weak decoherence condition 
\be
Re \, \dab \approx 0, \hspace{.15in}  \alpha \neq \beta 
\ee
and the medium decoherence condition 
\be
\dab \approx 0 , \hspace{.15in} \alpha \neq \beta.
\ee
Probability of a history $\alpha $ (in a decoherent family) is given by 
\be
p(\alpha) = d(\alpha, \alpha).
\ee
The assumed properties of decoherence functionals ensure that this probability 
assignment satisfies the usual probability sum rules.

\vspace{.15in}
\noindent
\textbf{5.2 Quantum mechanics of space-time} 

\hspace{.12in}
General relativity (GR) is a theory of gravity as well as of space-time 
geometry. 
It is supposed to be the unique low energy limit of any quantum theory of 
gravity [26,27]. In this subsection, our objective is to construct a 
generalized quantum mechanics of space-time which, in an appropriate limit, 
gives equations of GR (in a sense analogous to the way equations of classical 
mechanics were obtained from the history version of quantum mechanics in the 
previous section). 

The low energy theory has, as fine-grained histories, 4-dimensional manifolds 
equipped with smooth metrics of Lorentzian signature satisfying the Einstein 
equation and the matter field configurations satisfying appropriate field 
equations. For our quantum theory, therefore, we take as fine-grained  
histories 4-manifolds with arbitrary (continuous but not necessarily 
differentiable) Lorentz-signatured metrics $g_{\mu \nu}(x)$ and arbitrary 
(continuous) matter field configurations $ \phi (x)$. 

Generalized quantum mechanics permits topology change in space-time during 
quantum evolution. However, for simplicity, we shall consider only spacetimes 
with the fixed topology $ I \times M^3$ where I is a finite interval on the real 
line and $ M^3$ a closed (i.e. compact without boundary) 3-manifold (which 
correspond to  spatially compact  universes over  finite cosmological 
time-intervals). Accordingly, the 4-manifolds of interest have two $M^3$ 
boundaries; we shall call them $\partial M^{\prime}$ and 
$\partial M^{\prime \prime}$. The induced 3-metrics and 3-dimensional field 
configurations on these boundaries will be denoted as \\
$(h_{ij}^{\prime}(\textbf{x}), \chi^{\prime}(\textbf{x})) $ and 
$ (h_{ij}^{\prime \prime}(\textbf{x}), \chi^{\prime \prime}(\textbf{x}))$ 
respectively.

In a gauge in which $ g_{00}(x) = -1$ and $ g_{0j} =0$ ( the so - called 
Gaussian gauge), we can write the 4-dimensional line element as 
\be
ds^2 = -dt^2  + h_{ij}(\textbf{x},t)dx^i dx^j. 
\ee
The functions $ h_{ij}(\textbf{x},t)$ describe a family of 3-metrics 
parametrised by the proper time t. In this description, therefore, the 
fine-grained histories may be thought of as curves in a space (called 
\emph{superspace}) a typical  point of which jointly represents a  3-metric 
$h_{ij}(\textbf{x})$ and a spatial matter field configuration 
$ \chi (\textbf{x})$. 

Allowed coarse-grainings in the generalized quantum mechanics being 
developed are partitions of the fine-grained histories into 
(exclusive, exhaustive) diffeomorphism-invariant classes. The most 
general description of such a coarse-graining is given in terms of ranges of 
values of diffeomorphism-invariant functionals of the 4-geomety and matter 
field configurations.

Statements about the universe in terms of observable features like  
near-homogeneity and isotropy of 3-geometry at late enough times 
generally corresponds to partitioning of its histories into two 
diffeomorphism-invariant classes--one in which the statement is 
true and the other  in which it is not. (These classes are diffeomorphism- 
invariant because the statement does not involve coordinates or any other 
diffeomorphism non-invariant quantity.) 

The so-called class operators $C_{\alpha}$ and the decoherence functionals 
$ \dab $ are constructed in ref [25] essentially along the lines of the path 
integral constructions in eqs (31) and (32) taking due care of the 
invariances of the gravitational action. We shall give only a very rough 
outline of the constructions. 

The Hilbert space on which the action of the operators $\ca$ is defined is 
formally the space $ \mathcal{H}^{(h, \chi)}$ of square-integrable 
functionals of 3-metrics and matter field configurations on the boundary 
hypersurfaces $\partial M^{\prime}$ and $\partial M^{\prime \prime}$ 
mentioned earlier. (There are two Hilbert spaces involved; they are 
supposedly isomorphic.) 

The expression for the matrix elements of $\ca$ between the `coordinate 
eigenstates' (analogous to $|\qp>$ and $|\qdp>$) arrived at, in ref[23], 
after considerable groundwork relating to the Hamiltonian treatment of the 
relevant gauge-invariances (diffeomorphisms and their phase space extensions) 
and their quantum treatment along the traditional Faddeev-Popov (FP) lines, is 

\be
<h^{\prime \prime},\chi^{\prime \prime}|\ca | h^{\prime}, \chi^{\prime}>  =
 \int _{\alpha} Dg D\phi \delta [\Phi[g, \phi]] \Delta_{\Phi}[g, \phi] 
e^{iS[g, \phi]}. 
\ee 

\noindent
Here the functional integrations are over the 4-dimensional metrics g and the 
field configurations $\phi$ interpolating between the the boundary data 
appearing on the left. The $\delta$-functional represents the gauge condition 
employed and $\Delta_{\Phi}$ the corresponding FP factor; $S[g, \phi]$ is 
the 4-dimensional gravitational plus matter field action.

The constructions in ref [25] employ a time-symmetric formulation of the 
history formalism [28] which involves both the initial and the final density 
operators. (The traditional formalism is recovered as a special case of this by 
replacing the final density operator by the identity operator.) The initial 
and final density operators are assumed to be of the form 
\be
\rho_{in} = \sum_i p_i^{\prime}|\Psi_i><\Psi_i| \hspace{.62in} 
\rho_{f} = \sum_j p_j^{\prime \prime} |\Phi_j><\Phi_j|. 
\ee.
The decoherence functional is proposed to be given by 
\be
\dab &  = & \mathcal{N}Tr (\rho_f \ca \rho_{in} \cbd ) \\ \nonumber 
     &  = & \mathcal{N} \sum_{i,j} p_j^{\prime \prime} <\Phi_j | \ca |\Psi_i> 
     <\Phi_j | \cb |\Psi_i >^* p_i^{\prime} 
\ee 
where $\mathcal{N}$ is a constant ensuring correct normalisation of the 
decoherence functional. The Hilbert space matrix elements of the class 
operators appearing in eq(65) are related to the objects in eq(63) by the 
relation 
\be
< \Phi_j | \ca | \Psi_i> = \Phi_j [h^{\prime \prime}, \chi^{\prime \prime}] 
\circ <h^{\prime \prime}, \chi^{\prime \prime}| \ca | h^{\prime}, \chi^{\prime}> \circ \Psi_i [h^{\prime}, \chi^{\prime}] 
\ee
where $ \circ $ is a Hermitian inner product between functionals on the 
superspace 
(not necessarily a positive definite one). Eq(66) is analogous to the relation 
(valid in a Hilbert space of square-integrable functions )
\be 
< \phi | A | \psi > = \int \phi (x)^* <x | A | y > \psi (y) dx dy. 
\ee
[Note that, writing $<x|A|y> = f_y(x)$, the x-integration in eq(67) (and 
similarly the y-integration) inolves a scalar product. These scalar products 
are 
generally positive definite in the traditional formalism of quantum mechanics. 
In the formalism under consideration, however, this condition of positive 
definiteness can be dispensed with because positive definiteness of the 
probabilities in the theory is guaranteed by the defining properties of 
decoherent functionals. For the scalar product $\circ$ in eq(66), Hartle 
settles on the DeWitt inner product given on page(195)of ref[25].]

The wave functions $\Phi_j$ and $\Psi_i$ employed above have no direct 
probability interpretation; they appear only as part of the specification 
of the decohernce functional (65) ( which is to perform its ususual job of 
determining which families of coarse-grained histories are decoherent and to 
assign probabilities to histories in a decoherent family). 

The formalism presented above is generally covariant and circumvents the problem of time in quantum gravity. We did not need any preferred sets of space-like 
hypersurfaces in in our constructions. There is no notion of state on a 
spacelike hypersurface. 

In traditional quantum mechanics (with a given background spacetime) we do 
have a notion of state on a spacelike hypersurface. The formalism presented 
above, therefore, has the obligation to show how such a notion of state is 
recovered in appropriate limit. 

This part of the program has not been completed in ref [25] (nor in any later 
work as far as the author is aware). Some useful points in this connection have been made in ref [25]; we shall, however, skip them. 
    
\vspace{.2in}
\noindent
\textbf{VI. Some Mathematical Developments in Histories-based Theories}

\vspace{.12in}
In this section, we shall briefly describe some works aimed at an organized 
mathematical development of histories-based theories building up on the idea 
of Gell-Mann and Hartle of taking histories as the basic objects and enriching 
the formalism with additional concepts like a generalization of the concept of 
time sequences based on partial semigroups, a multi-time generalization of 
quantum logic (to describe histories as propositions) etc. 

\vspace{.12in}
\noindent 
\textbf{6.1 Generalization of the concept of time-sequences; partial 
semigroups} 

\vspace{.12in}
In histories-based theories, the time-sequences employed in the description 
of histories like 
\be
\alpha = (\alpha_{t_1}, \alpha_{t_2},...,\alpha_{t_n}) \hspace{.15in} 
t_1 < t_2 < ... < t_n 
\ee 
(where $\alpha_{t_i}$ are Schr$\ddot{o}$dinger picture projection operators) 
serve only for book-keeping; the properties of time t as a real variable are 
not used. The mathematical structure which correctly describes the 
book-keeping of this sort and also makes provision for useful generalizations of the concept of time-sequence is that of a partial semigroup[11]. 

A \emph{partial semigroup} (psg) is a nonempty set $\mathcal{K}$ (whose 
elements 
will be denoted as s,t,u,...) in which a binary operation $ \circ $ between 
certain pairs of elements is defined such that $ (s \circ t) \circ u = 
s \circ (t \circ u) $ whenever both sides are defined. A homomorphism of a psg 
$\mathcal{K}$ into another psg $ \mathcal{K}^{\prime}$ is a mapping 
$ \sigma : \mathcal{K} \rightarrow \mathcal{K}^{\prime}$ such that, for all 
$s,t \in \mathcal{K} $ with $s \circ t$ defined, $ \sigma (s) \circ \sigma (t)$ is also defined  and 
\be 
\sigma(s \circ t) = \sigma(s) \circ \sigma(t). 
\ee
If $\sigma$ is inverible, it is called an isomorphism (automorphism if 
$ \mathcal{K}^{\prime} = \mathcal{K}$).

The psg involved in the book-keeping of temporal supports of histories in 
traditional quantum mechanics is 
\begin{eqnarray*} 
\mathcal{K}_1 = \{ \textnormal{finite ordered subsets of R} \} 
\end{eqnarray*} 
whose general element is of the form 
\be
t = \{ t_1,t_2,...,t_n \}; \hspace{.12in} t_1 < t_2 < ... < t_n. 
\ee
If $ s = \{ s_1,...,s_m \} \in \mathcal{K}_1 $ such that $ s_m < t_1$, then 
$ s \circ t$ is defined and 
\be
s \circ t = \{ s_1,s_2,...,s_m,t_1,t_2,...,t_n \}. 
\ee 
It is useful to adopt the 
convention [25] that 
\be
\{ t_1 \} \circ \{ t_1 \} = \{ t_1 \}. 
\ee 
With this convention, we have $ s \circ t $ defined when $ s_m \leq t_1.$ Note 
that elements of $\mathcal{K}_1$ admit an irreducible decomposition of the 
form 
\be 
t = \{ t_1 \} \circ \{ t_2 \} \circ ... \circ \{ t_n \}. 
\ee 
Elements $ \{ t_i \} $ which cannot be further decomposed are called 
\emph{nuclear}.

Interesting examples of partial semigroups whose elements are defined in terms 
of light cones and which are useful in the construction of histories in 
curved spacetimes may be found in ref[11] and [2]. These types of 
constructions have the potential to contribute towards solving the problem 
of time in quantum gravity. 

\vspace{.15in} 
\noindent
\textbf{6.2 Quasitemporal structures [11,2]} 

\vspace{.12in} 
Histories of the form (68) also constitute a psg; we shall call it 
$\mathcal{K}_2$. For $ \alpha = (\alpha_{s_1},...,\alpha_{s_m})$ and 
$ \beta = (\beta_{t_1},...,\beta_{t_n})$ with $ s_m < t_1, \alpha \circ \beta $ is defined and is given by 
\be
\alpha \circ \beta = (\alpha_{s_1},...,\alpha_{s_m}, \beta_{t_1},...,\beta_{t_n}).
\ee 
There is a homomorphism $\sigma$ from $ \mathcal{K}_2$ onto $\mathcal{K}_1$ 
given by 
\be
\sigma(\alpha) = s, \hspace{.12in} \sigma(\beta) = t, \hspace{.12in} 
\sigma(\alpha \circ \beta) = s \circ t. 
\ee 
The triple ($\mathcal{K}_2, \mathcal{K}_1, \sigma $) defines a 
\emph{quasitemporal structure}( a pair of psgs with a homomorphism of one onto 
the other). This concept formalizes and generalizes the idea of histories as 
temporal sequences of `events'. 

Taking a clue from the traditional proposition calculus[30,31] where single 
time propositions are the basic entities, Isham[11]  suggested that histories 
must be treated as multitime or more general propositions. He evolved a scheme 
of \emph{quasitemporal theories} in which the basic objects were a triple 
$ (\mathcal{U}, \mathcal{T}, \sigma)$ defining a quasitemporal structure. 
The space $\mathcal{U}$ was called the \emph{space of history filters} and was 
assumed to be a meet semilattice with the operations of partial order $\leq$ 
(coarse-graining) and a meet/and operation $\wedge$ (simultaneous realization 
of two 
histories). The space $\mathcal{T}$ was called the \emph{space of temporal 
supports}. To accomodate the operation of negation of a history, he proposed 
that the space $\mathcal{U}$ be embedded in a larger space $\Omega$ (denoted as 
$\mathcal{U} \mathcal{P}$ in ref[11,32]) called the \emph{space of history 
propositions}. This larger space was envisaged as having a lattice structure. 
(Later, we shall see in section 6.4 that the proper mathematical structure for 
it is that of an orthoalgebra.) Decoherence functionals were defined as 
complex valued functions on the space $\Omega \times \Omega$ satisfying the usual 
four conditions of hermiticity, positivity, biadditivity and normalization.

\vspace{.15in}
\noindent
\textbf{6.3 The history projection operator(HPO) formalism[11,33]} 

\vspace{.12in} 
The logical operations on the history propositions are very similar to those 
of the traditional quantum logic[30,31] in which the elementary 
propositions (quantum events) have a standard representation 
as projection operators on the quantum- mechanical Hilbert space $\mathcal{H}$ 
of the system. It is natural to look for a similar representation for the 
history propositions. Now, the usual class operators $\ca$ occurring in the 
description of histories are products of (Heisenberg picture) projection 
operators. The product of two or more (generally noncommuting) projection 
operators need not be a projection operator. It follows that the class 
operators are generally not projection operators. 

It is useful to note in this connection that the tensor product $P \otimes Q$ 
of two projection operators P and Q on $\mathcal{H}$ is a projection operator 
on $\mathcal{H} \otimes \mathcal{H}$. Indeed, given (bounded) operators A,B,... on $\mathcal{H}$, we have, on $\mathcal{H}\otimes \mathcal{H}$, 
\begin{eqnarray*}
(A \otimes B)(C\otimes D) = AC \otimes BD; \hspace{.12in} 
(A \otimes B)^{\dagger} = A^{\dagger} \otimes B^{\dagger}.
\end{eqnarray*} 
This gives 
\begin{eqnarray*} 
(P \otimes Q)^2 = P^2 \otimes Q^2 = P \otimes Q; \hspace{.12in} 
(P \otimes Q)^{\dagger} = P^{\dagger} \otimes Q^{\dagger} = P \otimes Q. 
\end{eqnarray*}
It follows that histories of the form (68)can be represented as projection 
operators in the Hilbert space 
\be
\tilde{\mathcal{H}} = \mathcal{H}_{t_1}\otimes \mathcal{H}_{t_2} \otimes ... 
\mathcal{H}_{t_n} 
\ee
(where $ \mathcal{H}_{t_i} $ are copies of the Hilbert space $\mathcal{H}$) 
in the form 
\be
\alpha = \alpha_{t_1} \otimes \alpha_{t_2} \otimes ... \otimes \alpha_{t_n}.
\ee. 

Not all projection operators on $\tilde{\mathcal{H}}$ are `homogeneous' 
projection operators of the form (77). A sum of two or more mutually 
orthogonal projection operators is a projection operator. When such a sum is 
not expressible as a homogeneous projection operator, it is called an 
inhomogeneous projection operator and a history represented as such an 
operator is called an inhomogeneous history. 

The representation (77) of histories works well with the operation of negation. Denoting the negation of a history proposition $\alpha$ by $ \neg \alpha, $ 
we have, for $ \alpha = P \otimes Q$, 
\be
\neg \alpha & = & \neg (P \otimes Q) = I \otimes I - P \otimes Q \\ \nonumber 
            & = & (I-P) \otimes Q + P \otimes (I-Q) + 
	          (I-P) \otimes (I-Q) \\  \nonumber
	    & = & (\neg P) \otimes Q + P \otimes (\neg Q) + 
	          (\neg P) \otimes (\neg Q). 
\ee 
The right hand side is a sum of three mutually orthogonal homogeneous 
projection operators which represent the three mutually exclusive (and 
exhausive) situations corresponding to the negation of $ \alpha$. 

For some further developments in history theory employing the HPO formalism, 
see Griffiths [7]. 

\vspace{.12in}
\noindent 
\textbf{6.4 The algebraic scheme of Isham and Linden} 

\vspace{.12in}
In ref[32], Isham and Linden proposed a scheme more general than the 
quasitemporal formalism of Isham. In this scheme, even the concept of a 
quasitemporal structure was dispensed with. (The main motive for such a 
generalization comes from quantum gravity where one has the prospect of 
a `timeless' 
formulation of dynamics.) In this scheme, the basic ingredients are the  space 
$ \Omega$ (denoted as $\mathcal{U} \mathcal{P}$ in their work) of history 
propositions and the space $\mathcal{D}$ of decoherence functionals. The space 
$\Omega$ is assumed to be equipped with a structure (essentially that of an 
orthoalgebra[34]) incorporating the following operations :

\noindent
(i) Partial order ($\leq$). (a) Given $\alpha, \beta \in \Omega$, $\alpha 
\leq \beta$ means that $\alpha$ implies $\beta$ ( or is finer than $\beta$) 
or, equivalently, that $\beta$ is coarser than $\alpha$ (or a coarse-graining 
of $\alpha$). \\
(b) The space $\Omega $ has two distinguished elements: the 
\emph{null history proposition} 0 (a proposition that is always false) and 
a \emph{unit history proposition} 1 which is always true.
For all $\alpha \in \Omega$, we have 
\be
0 \leq  \alpha \leq 1. 
\ee

\noindent
(ii) Disjointness ($\perp$). This operation represents mutual exclusion: 
$\alpha \perp \beta$ means that if one of $\alpha$ and $\beta$ is realised, 
the other 
must be ruled out (they are \emph{disjoint}). 

\noindent
(iii) Disjoint join operation ($\oplus$). (a) Given $\alpha \perp \beta$ in 
$\Omega$, $\alpha \oplus \beta$ means `$\alpha $ or $ \beta$'. It is asumed 
to be  
commutative and associative :
\begin{eqnarray*}
\alpha \oplus \beta =  \beta \oplus \alpha; \hspace{.12in} 
(\alpha \oplus \beta) \oplus \gamma = \alpha \oplus (\beta \oplus \gamma) 
\end{eqnarray*}
whenever the expressions involved in these equations are defined. \\
(b) The operations $\leq$ and $\oplus$ are related by the requirement that 
$\alpha \leq \beta$ if and only if there exists an element $\gamma \in 
\Omega$ such that $ \beta = \alpha \oplus \gamma$. 

\noindent
(iii) negation ($\neg$). For every $ \alpha \in \Omega$, there is a unique 
element $ \neg \alpha$ in $\Omega$ (meaning `not $\alpha$' or negation of 
$\alpha$) such that $ \alpha \oplus \neg \alpha = 1.$ The negation operation 
satisfies the condition $ \neg (\neg \alpha) = \alpha$. (This follows from the 
defining conditions of $ \neg \alpha$ and the uniqueness of $\neg \alpha$.) 

\noindent
Note. The conditions defining an orthoalgebra are much weaker than those 
defining an (orthocomplemented) lattice. On a lattice, we have two connectives 
(binary operations $ \Omega \times \Omega \rightarrow \Omega$ ), $\wedge$ 
(meet: 
$alpha \wedge \beta $ means `$\alpha$ and $\beta$') and $\vee$ (join : 
$\alpha \vee \beta $ means `$\alpha$ or $\beta$') which are defined for every 
pair of elements. In contrast, an orthoalgebra has only one partial binary 
operation $\oplus$. 

The space $\mathcal{D}$ is treated as in section 6.2. The additivity condition 
on a decoherence functional now takes the following form: Given 
$\alpha \perp \beta$, we have 
\begin{eqnarray*}
d(\alpha \oplus \beta,\gamma) = d(\alpha, \gamma) + d(\beta, \gamma). 
\end{eqnarray*}

A set $\alpha^{(i)}$(i=1,...,n) of history propositions is \emph{exclusive} if, 
members of this set are pairwise disjoint; it is called \emph{exhaustive} 
(or \emph{complete}) if it is exclusive and 
$\alpha^{(1)} \oplus ... \oplus \alpha^{(n)} = 1.$ 
Consistency /decoherence conditions and probability assignments on complete 
sets of history propositions are done in terms of decoherence functionals as 
before.

\vspace{.15in}
\noindent
\textbf{6.5 An axiomatic scheme for quasitemporal histories-based theories; 
symmetries and conservation laws in histories-based theories }

\vspace{.12in}
This section contains a brief description of some work  by the author and 
Yogesh Joglekar [29,2]. The work presented in ref[29] is an axiomatic 
development of dynamics of 
systems in the framework of histories and contains history versions of 
classical and traditional quantum mechanics as special cases. This work was 
motivated by the observation that while histories of a system contain, in 
principle, information about its measurable properties and dynamics, the 
Isham-Linden formalism is not adequately equipped to bring it out in an 
autonomous   framework. We developed a more elaborate formalism for 
quasitemporal theories to fulfil this need.

Before stating the (five) axioms, we give a few more definitions relating to parial 
semigroups (psgs). A \emph{unit element} in a psg $\mathcal{K}$ is an element 
e such that $e \circ s = s \circ e = s$ for all $s \in \mathcal{K}$. An 
\emph{absorbing element} in $\mathcal{K}$ is an element a such that 
$ a \circ s = s \circ a = a $ for all $s \in \mathcal{K}$. A psg may or may 
not have a unit and/or absorbing element; when either of them exists, it is 
unique. In a psg, elements other than the unit and absorbing elements are 
called \emph{typical}. 

A psg is called \emph{directed} if, for any two typical elements s and t 
in it, if 
$s \circ t$ is defined, then $ t \circ s$ is not defined. The 
psgs $\mathcal{K}_1$ and $\mathcal{K}_2$ introduced in section 6.2 are 
directed. Directedness of the two psgs defining a quasi-temporal structure 
reflects the presence, in the quasi-temporal structure, a direction of flow 
of `time'.

The set of nuclear elements in a psg $\mathcal{K}$ will be denoted as 
$\mathcal{N}(\mathcal{K})$. Clearly, $ \mathcal{N}(\mathcal{K}_1)$ = R, 
the set of real 
numbers. A psg is called \emph{special} if its elements admit semi-infinite 
irreducible decompositions [modulo redundancies implied by the convention (72)].

\vspace{.12in}
\noindent
Axiom $A_1$ \emph{(Quasitemporal Structure Axiom)}. Associated with a (closed) 
dynamical system S is a \emph{history system} ($ \mathcal{U},\mathcal{T},
\sigma$) defining a quasitemporal structure as described earlier. The psgs 
$\mathcal{U}$ and $\mathcal{T}$ are assumed to be special and and to satisfy 
the relation $ \sigma [\mathcal{N}(\mathcal{U})] = \mathcal{N}(\mathcal{T})$. 

Elements of $\mathcal{U}$ (history filters)  are denoted as $\alpha, \beta,...$ and those of $\mathcal{T}$ (temporal supports) as $ \tau, \tau^{\prime},....$ 
If $\alpha \circ \beta $ and $\tau \circ \tau^{\prime}$ are defined, we write 
$\alpha \triangleleft \beta $ ( which means $\alpha$ precedes $\beta$) and  
$\tau \triangleleft \tau^{\prime}$.

\vspace{.12in}
\noindent
$A_2$ (Causality axiom). If $ \alpha, \beta,...,\gamma \in \mathcal{N}
(\mathcal{U})$ are such that $\alpha \triangleleft \beta \triangleleft ... 
\triangleleft \gamma$ with $\sigma(\alpha) = \sigma(\gamma)$, then we must have 
$\alpha = \beta = ...=\gamma$.

In essence, this axiom forbids histories corresponding to `closed time loops'. 
(This is the most primitive version of causality.) From these two axioms 
one can 
prove [29] that the two psgs $\mathcal{U}$ and $\mathcal{T}$ are directed and 
some other useful results. It is interesting to note that (in the present 
quasitemporal setting) a primitive version of causality implies a direction of 
flow of `time'.

\vspace{.12in}
\noindent
$A_3$ (Logic Structure Axiom). Every space $\mathcal{U}_{\tau} \equiv 
\sigma^{-1}(\tau)$ for a $\tau \in \mathcal{N}(\mathcal{T})$ has the structure 
of a logic as defined in Varadarajan's book [31].

A logic is essentially an orthoalgebra with meet ($\wedge$) and join($\vee$) 
operations defined in it. Ref[29] contains the full statement of the axiom 
$A_3$ detailing the logic structure. Note that isomorphism of 
$\mathcal{U}_{\tau}s$ (as logics) for different nuclear $\tau s$ is not assumed. This generality gives the 
formalism  additional flexibility so as to make it applicable to systems 
whose empirical characteristics may change with time ( for example, the 
universe).

Using the logic structure of $\mathcal{U}_{\tau}s$, a larger space $\Omega$---
the space of history propositions --- was explicitly constructed and shown to be 
an ortho-algebra as envisaged in the scheme of Isham and Linden. Its subspace 
$\tilde{\mathcal{U}}$ representing the `homogeneous histories' (which is 
obtained from $\mathcal{U}$ after removing some redundancies) was shown to be 
a meet-semilattice as envisaged in the quasitemporal sheme of Isham [11]. 

\vspace{.12in}
\noindent
$A_4$ (Temporal Evolution Axiom). The temporal evolution of a system as in 
axiom $A_1$  is given, for each pair $\tau, \tau^{\prime} \in 
\mathcal{N}(\mathcal{T})$ such that $\tau \triangleleft \tau^{\prime}$, by a 
set of mappings $V(\tau, \tau^{\prime})$ of $\mathcal{U}_{\tau}$ onto 
$\mathcal{U}_{\tau^{\prime}}$ which are logic homomorphisms (not necassarily 
injective) and which satisfy the composition rule 
$ V(\tau^{\prime \prime},\tau^{\prime}).V(\tau^{\prime}, \tau) = 
V(\tau^{\prime \prime},\tau)$ whenever $\tau \triangleleft 
\tau^{\prime} \triangleleft \tau^{\prime \prime}$ and 
$\tau \triangleleft \tau^{\prime \prime}.$ (Note that the relation 
$\triangleleft$ is generally not transitive.)

The space $\mathcal{D}$ of decoherence functionals in the Isham-Linden scheme 
is not properly integrated with the basic framework. It is, for example, not 
made clear as to what distinguishes different elements of this space and when 
is a particular element relevant. In traditional quantum mechanics, as we 
have seen, a decoherence functional is determined by the evolution operator 
and the initial state. In the Isham-Linden formalism, there is no explicitly 
defined concept of state or of evolution; information about both of these is 
contained in the decoherence functional. It is, however, not clear how to bring out and use this information.

In our formalism we have both the concepts defined. A state at `time' 
$\tau \in \mathcal{N}(\mathcal{T})$ is essentially a generalized probability 
on $\mathcal{U}_{\tau}$ [29]. Axiom $A_5$ associates a decoherence functional 
with a given law of temporal evolution and a given initial state, stipulates 
the (weak) decoherence condition and gives the usual rule for probability 
assignment for  consistent/decoherent histories (elements of $\Omega$). 

The formalism would be complete if an explicit expression for the decoherence 
functional (analogous to those in section 3 and 4) is given. We were able to do it only for the Hilbert space -based theories (which have the traditional 
quantum mechanics as a special case) and for classical mechanics. 

In the second paper [2], a systematic treatment of symmetries and 
conservation laws 
in the formalism of ref [29] was given. Symmetries were defined in a 
straightforward manner as natural invariances (automorphisms) of the basic 
structure. The directedness of the psg $\mathcal{T}$ leads to a 
classification of symmetries into orthochronous (those preserving the 
`temporal order' of events) and non-orthochronous.

A straightforward criterion for physical equivalence of histories was given: 

\emph{All histories related to each other through orthochronous symmetry 
operations are physically equivalent.}

\noindent
This criterion covers several different notions of physical equivalence of 
histories considered by Gell-Mann and Hartle [35] as special cases.

In familiar situations, a reciprocal relationship between traditional 
symmetries (Wigner symmetries in quantum mechanics and and Borel-measurable 
transformations  of phase space in classical mechanics) and symmetries 
in our formalism was established. 

In a somewhat restricted class of theories, definition of a conservation law 
was given in the history language which agrees with the standard ones in 
familiar situations. In a subclass of these theories, a Noether-type  theorem 
(implying a connection between continuous symmetries of dynamics and 
conservation laws) was proved.

The formalism evolved was applied to historis (of particles, fields, or 
more general objects) in general curved spacetimes. Sharpening the definition 
of symmetry so as to include a continuity argument, it was shown that a symmetry in our formalism implies a conformal isometry of the space-time metric. This 
condition is satisfied by all known symmetries in nature.

\vspace{.2in}
\noindent
\textbf{Acknowledgements}

The author would like to thank Professor R. Srinivasan, Professor N. Mukunda 
and the Centre For Learning for the hospitality at Vardanahalli.

\vspace{.12in}
\noindent
\textbf{References}
\begin{enumerate}
\item R.M.F. Houtappel, H. Van Dam and E.P. Wigner, Rev. Mod. Phys. 
\textbf{37}, 595 (1965).
\item Tulsi Dass and Y.N. Joglekar Ann. Phys. \textbf{287}, 191 (2001).
\item R.B. Griffiths, J. Stat. Phys. \textbf{36} 219 (1984).
\item R. Omn$\grave{e}$s, J. Stat. Phys. \textbf{53} 933 (1988).
\item R. Omn$\grave{e}$s, Rev. Mod. Phys. \textbf{64}, 339 (1992).
\item R. Omn$\grave{e}$s, `The Interpretation of Quantum Mechanics', Princeton 
University Press (1994).
\item R.B. Griffiths, Phys. Rev. \textbf{A54}, 2759 (1996).
\item R.B. Griffiths and R. Omn$\grave{e}$s, Physics Today (Aug 1999) p.26.
\item R. Omn$\grave{e}$s, `Understanding Quantum Mechanics', Princeton 
University Press (1999). 
\item R.B. Griffiths, `Consistent Quantum Theory', Cambridge University 
Press (2002). 
\item C.J. Isham, J. Math. Phys. \textbf{35}, 2157 (1994). 
\item R.B. Griffiths, Phys. Rev. \textbf{A66}, 062101 (2002). 
\item M.Gell-Mann and J.B. Hartle, Arxiv gr-qc/9509054 (1995). 
\item M.Gell-Mann and J.B. Hartle, in `Complexity, Entropy and the Physics of 
Information' (W. Zurek, Ed.), SFI Studies in the Science of Complexity, 
vol. 8, p.425, Addison-Wesley (1990). 
\item M. Gell-Mann and J.B. Hartle, Phys. Rev. \textbf{D47}, 3345 (1993). 
\item R.P. Feynman and F. L. Vernon, Jr., Ann. Phys., \textbf{24}, 118 (1963). 
\item A.O. Caldeira and A.J. Leggett, Physica \textbf{121}, 587 (1983). 
\item W. Zurek, Rev. Mod. Phys. \textbf{75}, 715 (2003). 
\item D. Giulini et al ,`Decoherence and the Appearance of the Classical 
World in Quantum Theory',Springer (1996). 
\item T.A. Brun, I.C. Percival and R. Schack, arxiv quant-ph/9509015 (1995). 
\item R.P. Feynman and A.R. Hibbs, `Quantum Mechanics and Path Integrals', 
McGraw-Hill (1965).
\item C. Kiefer, `Consistent Histories and Decoherence' in ref [19] (1996). 
\item T. Brun, Phys. Rev. \textbf{D47}, 3383 (1993). 
\item J.B. Hartle, in `Quantum Cosmology and Baby Universes' (S. Coleman, 
J. Hartle, T. Piran and S. Weinberg, Eds.), World Scientific (1991). 
\item J.B. Hartle, in `Gravitation and Quantization, Proceedings of the 1992 
Les Houches Summer School' (B. Julia andJ.Zinn-Justin, Eds.), Les Houches 
Summer School Proceedings, vol. 57, North Holland (1995). 
\item S. Deser, J.Grav. and Rel. \textbf{1},9 (1970). 
\item D. Boulware and S. Deser, Ann. Phys. \textbf{89}, 193 (1975). 
\item Y. Aharonov, P.G. Bergman and J.L. Lebowitz, Phys. Rev. \textbf{134}, 
1410B, (WZ, p.680) (1964). 
\item Tulsi Dass and Y. Joglekar, J. Math. Phys. \textbf{39}, 704 (1998); 
erratum, J. Math. Phys. \textbf{40}, 3235 (1999). 
\item J.M. Jauch, `Foundations of Quantum Mechanics', Addison-Wesley (1973). 
\item V.S. Varadarajan, `Geometry of Quantum Theory' 2nd ed., Springer (1985). 
\item C.J. Isham and N. Linden, J. Math. Phys. \textbf{35}, 5452 (1994). 
\item C.J. Isham, arxiv quant-ph/9506028 (1995). 
\item D.J. Foulis, R.J. Greechie and G.T. R$\ddot{u}$ttiman,Int. J. Theor. 
Phys. \textbf{31}, 789 (1992). 
\item M. Gell-Mann and J.B. Hartle, arxiv gr-qc/9404013 v3, (1996). 

\end{enumerate}

\end{document}